\documentclass[11pt]{article}
\usepackage{graphicx}

\newcommand{\BABARPubYear}    {02}

\newcommand{\BABARConfNumber} {20}
\newcommand{\SLACPubNumber} {9321}

\input pubboard/babarsym

\def\BDstarDsstar{\ensuremath{  B^{0} \rightarrow {  D_s^{*+} D^{*-}}}\xspace}
\def\BDstarDs{\ensuremath{  B^{0} \rightarrow {   D_s^{+} D^{*-}}}\xspace}
\def\BDstarDss{\ensuremath{  B^{0} \rightarrow {  D_s^{(*)+} D^{*-}}}\xspace}

\def\inter{i}

\def\Dspi{{\ensuremath{\Dsp\pi^-}}\xspace}

\def\Dsstarpi{{\ensuremath{\Dspstar\pi^-}}\xspace}

\def\DsspiWS{{\ensuremath{\Dsps\pi^+}}\xspace}

\def\brphipi{\ensuremath{\BR(\dstophipi)}\xspace}
\def\systphipi{\ensuremath{(\mathrm{syst}~\brphipi})\xspace}
\def\mmiss{\ensuremath{M_{\rm miss}}\xspace}
\def\Dsp{{\ensuremath{ D_s^+}}\xspace}

\def\Dspstar{{\ensuremath{ D_s^{*+}}}\xspace}
\def\Dsps{{\ensuremath{ D_s^{(*)+}}}\xspace}

\def\Dsphipi{{ \ensuremath{ {D_s^{+}}\rightarrow \phi \pi^{+}}}\xspace}
\def\Dsgamma{{ \ensuremath{ {D_s^{*+}}\rightarrow {D_s^{+}} \gamma}}\xspace}

\def\dstophipi{{\ensuremath{\Ds\to \phi\pi^{+}}}\xspace}
\def\dstokstark{{\ensuremath{\Ds\to \Kstarzb K^+}}\xspace}
\def\dstoksk{{\ensuremath{\Ds\to \KS K^+}}\xspace}

\def\phikk{{\ensuremath{\phi\to { K^+K^-}}}}

\long\def\inst#1{\par\nobreak\kern 4pt\nobreak
    {\it #1}\par\vskip 10pt plus 3pt minus 3pt}

\begin{document}
{\pagestyle{empty}

\begin{flushright}
\babar-CONF-\BABARPubYear/\BABARConfNumber \\
SLAC-PUB-\SLACPubNumber \\
July 2002 \\
\end{flushright}

\par\vskip 5cm

\begin{center}
\Large \bf \boldmath 
Measurement of \BDstarDss Branching Fractions 
and Polarization in the Decay \BDstarDsstar
with a Partial Reconstruction Technique
\end{center}
\bigskip

\begin{center}
\large The \babar\ Collaboration\\
\mbox{ }\\
{July 25, 2002}\\
\end{center}
\bigskip \bigskip

\begin{center}
\large \bf Abstract
\end{center}
We present measurements of the decays \BDstarDss, using data
recorded by the \babar\ detector in 1999 and 2000, consisting of
20.8\invfb. The analysis is conducted with a partial reconstruction
technique, in which only the \Dsps and the soft pion from the $D^{*-}$ 
decay are reconstructed.
From the observed rates, we measure the branching fractions 
${\cal B}(\BDstarDs) = (1.03 \pm 0.14 \pm 0.13 \pm 0.26) \%$
and 
${\cal B}(\BDstarDsstar) = (1.97 \pm 0.15 \pm 0.30 \pm 0.49) \%$,
where the first error is statistical, the second is systematic,
and the third is the error due to the \Dsphipi branching fraction uncertainty.
From the \BDstarDsstar angular distributions,  
we measure the fraction of longitudinal
polarization  
$\Gamma_L/\Gamma = (51.9\pm 5.0 \pm 2.8)\%$,
which is consistent with the theoretical prediction, based on
factorization. These results are preliminary.

\vfill
\begin{center}
Contributed to the 31$^{st}$ International Conference on High Energy Physics,\\ 
7/24---7/31/2002, Amsterdam, The Netherlands
\end{center}

\vspace{1.0cm}
\begin{center}
{\em Stanford Linear Accelerator Center, Stanford University, 
Stanford, CA 94309} \\ \vspace{0.1cm}\hrule\vspace{0.1cm}
Work supported in part by Department of Energy contract DE-AC03-76SF00515.
\end{center}

\newpage
} 

\begin{center}
\small

The \babar\ Collaboration,
\bigskip

B.~Aubert,
D.~Boutigny,
J.-M.~Gaillard,
A.~Hicheur,
Y.~Karyotakis,
J.~P.~Lees,
P.~Robbe,
V.~Tisserand,
A.~Zghiche
\inst{Laboratoire de Physique des Particules, F-74941 Annecy-le-Vieux, France }
A.~Palano,
A.~Pompili
\inst{Universit\`a di Bari, Dipartimento di Fisica and INFN, I-70126 Bari, Italy }
% ...

\end{center}\newpage

\section{Introduction}
\label{sec:Introduction}

Precise knowledge of the branching fractions of exclusive \B decay
modes provides a test of the factorization approach used for the
calculation of these branching fractions. Further tests are provided
by measuring the polarization in decays of \B mesons to vector-vector
final states.  Within current experimental sensitivities, these
measurements are consistent with the factorization predictions for the
final states $D^*\rho$~\cite{ref:cleo-dstrho},
$D^*\rho(1450)$~\cite{ref:dstrhoprime}, and
$D^*D_s^*$~\cite{ref:cleo-dds-polar}.

In this paper we present measurements of the branching
fractions\footnote[1]{Reference to a specific decay
channel or state also implies the charge conjugate decay or state. 
The notation \Dsps refers to either \Ds or \Dspstar.} ${\cal B}(\BDstarDss)$. We also report the measurement of
the \Dspstar polarization in the decay \BDstarDsstar, obtained from
an angular analysis.
These results provide increased precision tests of factorization.

\section{The \babar\ Detector and Dataset}
\label{sec:babar}
The data used in this analysis were collected with the \babar\
detector at the \pep2\ storage ring. An integrated luminosity of
20.8\invfb was recorded in 1999 and 2000 at the \FourS resonance,
corresponding to about 22.7 million produced \BB pairs.

Since a detailed description of the \babar\ detector is presented in
Ref.~\cite{ref:babar}, only the components of the detector most
crucial to this analysis are briefly summarized below.
Charged particles are reconstructed with a five-layer, double-sided
silicon vertex tracker (SVT) and a 40-layer drift chamber (DCH) with a
helium-based gas mixture, placed in a 1.5 T solenoidal field produced
by a superconducting magnet. The charged particle momentum resolution
is approximately $(\delta p_T/p_T)^2 = (0.0013 \, p_T)^2 +
(0.0045)^2$, where $p_T$ is given in \gevc.  The SVT, with a typical
single-hit resolution of 10\mum, provides measurement of the impact
parameters of charged particle tracks in both the plane transverse to
the beam direction and along the beam.
Charged particle types are identified from the ionization energy loss
(\dedx) measured in the DCH and SVT, and the Cherenkov radiation
detected in a ring imaging Cherenkov device (DIRC).  Photons are
identified by a CsI(Tl) electromagnetic calorimeter (EMC) with an
energy resolution $\sigma(E)/E = 0.023\cdot(E/{\rm GeV})^{-1/4}\oplus
0.019$.

\section{Method of Partial Reconstruction}
\label{sec:Analysis}

\par In reconstructing the decays \BDstarDss, with $ D^{*-}
\rightarrow \Dzb \pi^- $, no attempt is made to reconstruct the
\Dzb decays.  Only the \Dsps and the soft $\pi^-$ from the $
D^{*-}$ decay are detected. In this way, the candidate selection efficiency is 
higher by almost an order of magnitude than when performing 
full reconstruction of the final state. 
Given the four-momenta of the \Dsps and $\pi^-$ and assuming that
their origin is a \BDstarDss decay, the four-momentum of the \Bz may
be calculated up to an azimuthal angle $\phi$ about the \Dsps flight
direction. This calculation also makes use of 
the total beam energy in the center-of-mass (CM) system
and the masses of the \Bz and $D^{*-}$. 
Energy and momentum conservation then allows the calculation of the
four-momentum of the \Dzb, whose square yields the 
$\phi$-dependent missing mass
\begin{equation}
\mmiss = \sqrt { (E_{\rm beam} - E_{\Dsps }  - E_{\pi} )^2 
- ( {\bf p}_B - {\bf p}_{\Dsps }- {\bf p}_{\pi})^2  }. 
\end{equation}
In this analysis the missing mass is defined using an arbitrary choice
for the angle $\phi$, such that the \Bz momentum ${\bf p}_B$ makes the
smallest possible angle with ${\bf p}_{\pi}$ and ${\bf p}_{\Dsps}$ in the 
CM frame.

\section{\boldmath Event Selection}
\label{sec:selection}
For each event, we calculate the ratio of the second to the zeroth
order Fox-Wolfram moment~\cite{ref:FoxWolf}, using 
all charged tracks and neutral clusters in the event. This ratio is 
required to be less than 0.35 in order to 
suppress continuum $\epem \to \qqbar$ events, where $q = u,d,s,c$.

\subsection{\boldmath \Ds and \Dspstar Candidate Selection}
\label{sec:selDsps}

We reconstruct \Ds mesons in the decay modes \dstophipi,
\dstokstark and \dstoksk, with subsequent decays \phikk, $\Kstarzb\to
K^-\pi^+$ and $\KS\to\pi^+\pi^-$. These modes were selected since they
offer the best combination of branching fraction, detection efficiency
and signal-to-background ratio.  The charged tracks are required to
originate from within $\pm$10~cm along the beam direction and
$\pm$1.5~cm in the transverse plane, and leave at least 12 hits in the
DCH.

Kaons are identified using \dedx information from the SVT and DCH, and
the Cherenkov angle and the number of photons measured with the DIRC.
For each detector component $d = \{\rm SVT, \ DCH, \ DIRC\}$, a
likelihood $L^K_d$ ($L^\pi_d$) is calculated given the kaon (pion)
mass hypothesis.  A charged particle is classified as a ``loose'' kaon
if it satisfies $L^K_d / L^\pi_d > 1$ for at least one of the detector
components. A ``tight'' kaon classification is made if the condition 
$\prod_d L^K_d / L^\pi_d > 1$ is satisfied.

\begin{figure}
\begin{center}
        \includegraphics[width=0.75\textwidth]{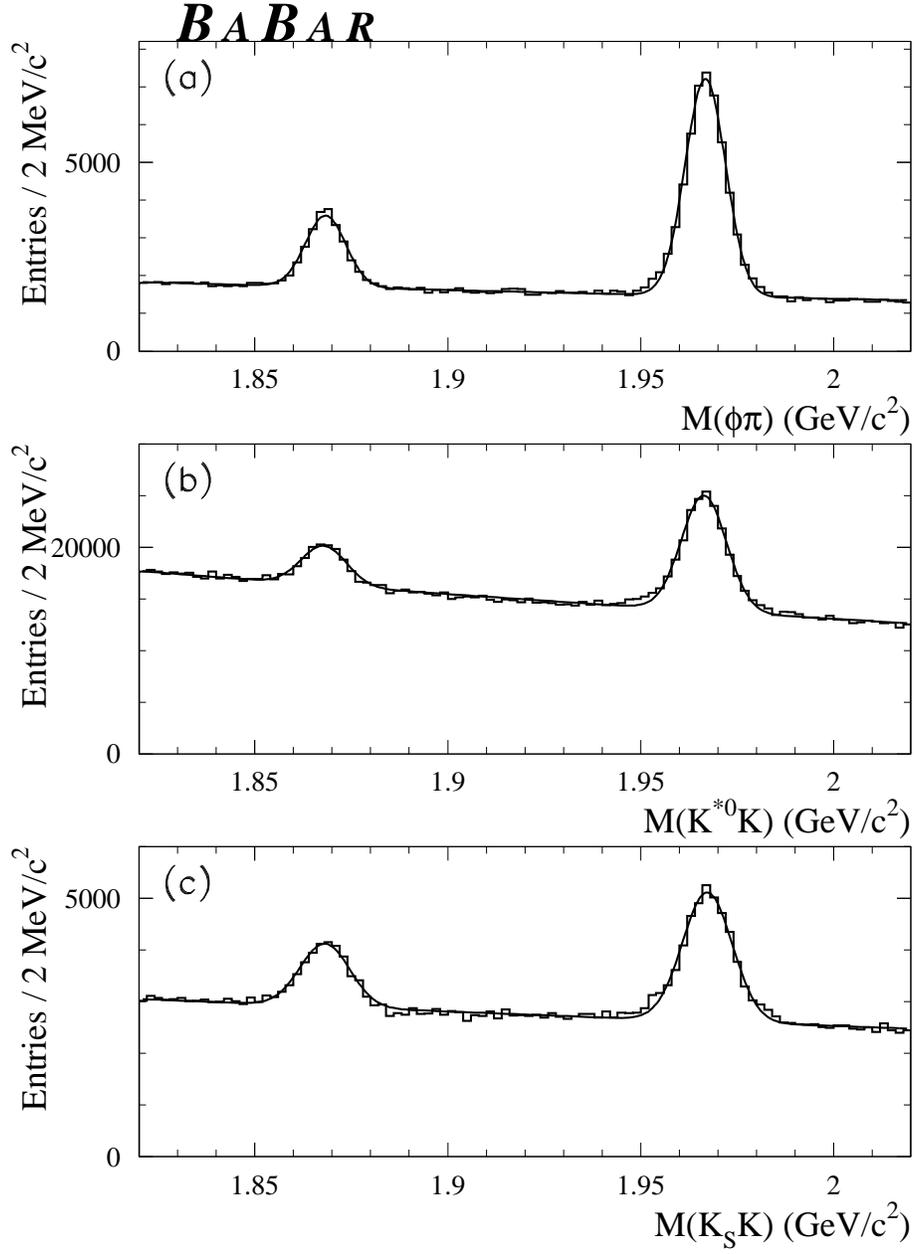} 
	\caption{The invariant mass
		spectra for (a) $\phi\pi^+$, (b) $\Kstarzb K^{+}$, (c) $\KS K^{+}$. 
		The peaks are due to the \Dp (left) 
		and \Ds (right) decays in this mode.
		The fit function is a single Gaussian for each peak, with
		the widths constrained to be equal, plus an exponential background.}
	\label{fig:ds_modes}
\end{center}
\end{figure}

Three charged tracks consistent with originating from a common vertex are
combined to form a \Ds candidate. 
In the case of the decay \Dsphipi, two oppositely charged tracks must
be identified as kaons with the loose criterion, with at least one
of them also satisfying the tight criterion.  No identification
criteria are applied to the pion from the \Ds decay.  The
reconstructed invariant mass of the ${K^+K^-}$ candidates must be
within 8\mevcc of the nominal $\phi$ mass~\cite{ref:pdg}.  In the
decay \Dsphipi, the $\phi$ meson is polarized longitudinally,
resulting in the kaons having a $\cos^2\theta_{H}$ distribution, where
$\theta_{H}$ is the angle between the $K^+$ and \Ds in the $\phi$ rest
frame.  We require $|\cos\theta_{H}|>0.3$, which retains 97\% of the
signal while rejecting about 30\% of the background. With these
requirements, the signal decay \Dsphipi and the Cabibbo-suppressed
decay $\Dp \rightarrow \phi \pip$ are readily observed
(Fig.~\ref{fig:ds_modes}a).

In the reconstruction of the \dstokstark mode, the $K^{-}\pi^{+}$
invariant mass is required to be within 65\mevcc of the central
$\Kstarzb$ mass~\cite{ref:pdg}. This wider window leads to a fraction
of combinatorial background much larger than in the \Dsphipi mode.  To
reduce this background, we require
$|\cos\theta_{H}|>0.5$. In addition, substantial background arises
from the decays $D^{+}\to \Kstarzb\pi^{+}$ and $D^{+}\to \Kzb\pi^{+}$,
which tends to peak around the nominal \Ds mass. This background is
suppressed by requiring that the kaon daughter of the $\Kstarzb$
satisfy the loose kaon identification criterion, and that the other
kaon satisfy the tight criterion.  Fig.~\ref{fig:ds_modes}b shows the
reconstructed $\Kstarzb K^+$ invariant mass.

For the decay mode \dstoksk, $\KS\to \pi^{+}\pi^{-}$, the
$\pi^+\pi^-$ invariant mass must be within 15\mevcc of the nominal
\KS\ mass, and the charged kaon is identified using the tight
criterion. To improve the purity of the \KS sample, we determine the
angle $\alpha$ between the \KS momentum and the flight direction defined by
its decay vertex and the primary vertex of the event. We require
$\cos\alpha>0.98$ to reject the combinatorial background.  
The $\KS K^+$ invariant mass distribution is shown in
Fig.~\ref{fig:ds_modes}c.

The invariant mass $M_{D_s}$ of all \Dsp candidate is required to be
within three standard deviations ($\sigma_{M_{D_s}}$) of the signal
distribution peak $M_{D_s}^{\rm peak}$ seen in the data.  The standard
deviations are $\sigma=5.23$\mevcc for $\phi\pi^+$, 5.97\mevcc for
$\Kstarzb K^+$, and 6.46\mevcc for $\KS K^+$.

\begin{figure}
\begin{center}
	\includegraphics[width=0.47\textwidth]{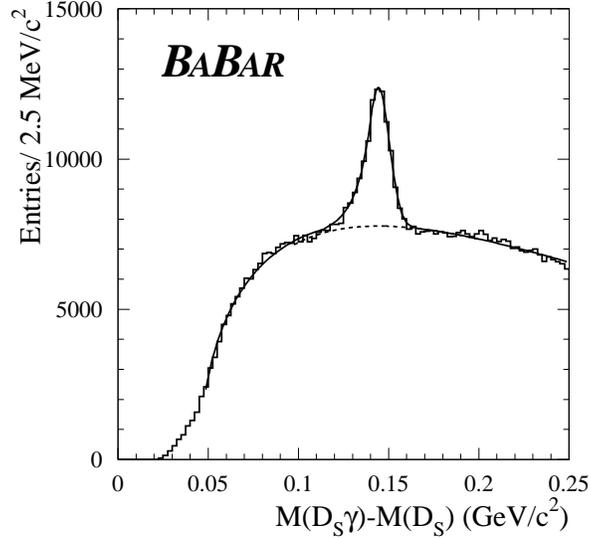}
\caption{ Distribution of the mass
difference $\Delta M = M_{\Ds\gamma} - M_{\Ds}$. All the \Ds decay modes
have been combined in this plot.  The fit function is a Crystal Ball
function for the signal plus a threshold function, as described in the
text.} 
\label{fig:dsst}
\end{center}
\end{figure}

All \Ds candidates satisfying all the above selection criteria are combined with
photon candidates to form \Dsgamma candidates. 
The candidate photons are required to satisfy $E_\gamma > 50$\mev,
where $E_{\gamma}$ is the photon energy in the laboratory frame, and
$E_\gamma^* > 110$\mev, where $E^*_{\gamma}$ is the photon energy in
the CM frame. When the photon candidate is combined with any other
photon candidate in the event, the pair must not form a good $\piz$
candidate, defined by a total CM energy $E_{\gamma\gamma}^* > 200$\mev
and an invariant mass $115 < M_{\gamma\gamma} < 155$\mevcc.

The distribution of the mass difference $\Delta M = M(\Ds
\gamma)-M(\Ds)$ of events satisfying these criteria is shown in
Fig.~\ref{fig:dsst}. The distribution of signal events is
parameterized with a Crystal Ball function~\cite{CBfunc}, which
incorporates a Gaussian core with a power-law tail toward lower
masses, and accounts for calorimeter shower shape fluctuations and
energy leakage.  The background is modeled by a threshold
function~\cite{ref:thresh}.

\subsection{\boldmath Selection of \BDstarDss Decays}
\label{sec:BSel}

\Dspstar candidates used in the partial reconstruction of the decay
\BDstarDsstar must satisfy $|\Delta M - \Delta M^{\rm peak}| < 2.5\,
\sigma_{\Delta M}$, where $\Delta M^{\rm peak}$ is the peak of the
signal $\Delta M$ distribution observed in the data, and
$\sigma_{\Delta M} = 5.7 \pm 0.3$\mevcc is its r.m.s.
The CM momentum of the \Dsps candidate
is required to be greater than 1.5\gevc.
\Dsps candidates satisfying these criteria, in addition to those
described in section~\ref{sec:selDsps}, are then combined with $\pi^-$
candidates to form partially reconstructed \BDstarDss candidates.

Due to the high combinatorial background in the $\Delta M$
distribution, more than one $\Dspstar\pi^-$ candidate pair per event
is found in 20\% of the events. To select the best candidate in the
event, the following $\chi^2$
\begin{equation}
\chi^2=
  \biggl(\frac{M_{\inter}-M_{\inter}^{\rm peak}}{\sigma_{\inter}}\biggr)^2+
  \biggl(\frac{M_{D_s}-M_{D_s}^{\rm peak}}{\sigma_{D_s}}\biggr)^2+
  \biggl(\frac{\Delta M-\Delta M^{\rm peak}}{\sigma_{\Delta m}}\biggr)^2
\end{equation}
is calculated for each \Dspstar candidate, where $M_{\inter}$ is
the invariant mass of the intermediate $\phi$, $K^{*0}$, or \KS candidate,
depending on the \Dsp decay mode, $M_{\inter}^{\rm peak}$ is the
corresponding peak of the signal $M_{\inter}$ distribution, and
$\sigma_{\inter}$ is its width. Only the candidate with the
smallest value of $\chi^2$ in the event is accepted. 

\section{Results}
\label{sec:results}

\begin{figure}
\begin{center}
        \includegraphics[width=0.9\textwidth]{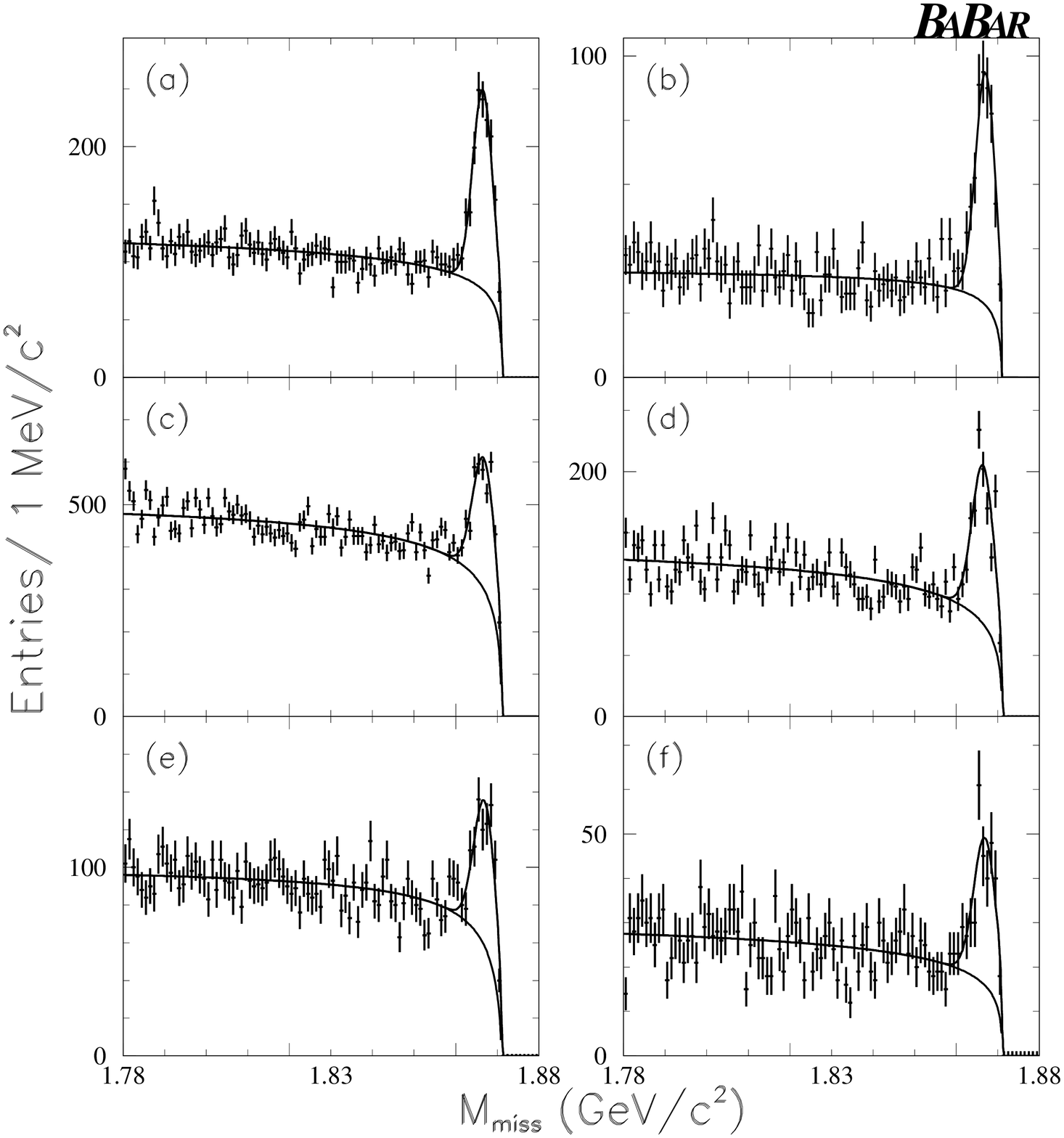} 
	\caption{Missing mass distributions of data events.
		(a) $\Dspi$ with \Dsphipi, 
		(b) $\Dsstarpi$ with   \Dsphipi,
		(c) $\Dspi$ with \dstokstark,
		(d) $\Dsstarpi$ with \dstokstark,
		(e) $\Dspi$ with \dstoksk, 
		(f) $\Dsstarpi$ with \dstoksk.
		 The curves show the result of the fit (see text), 
		 indicating the signal 
		 and background contributions.}
	\label{fig:signal}
\end{center}
\end{figure}

\begin{figure}
	\includegraphics[width=0.47\textwidth]{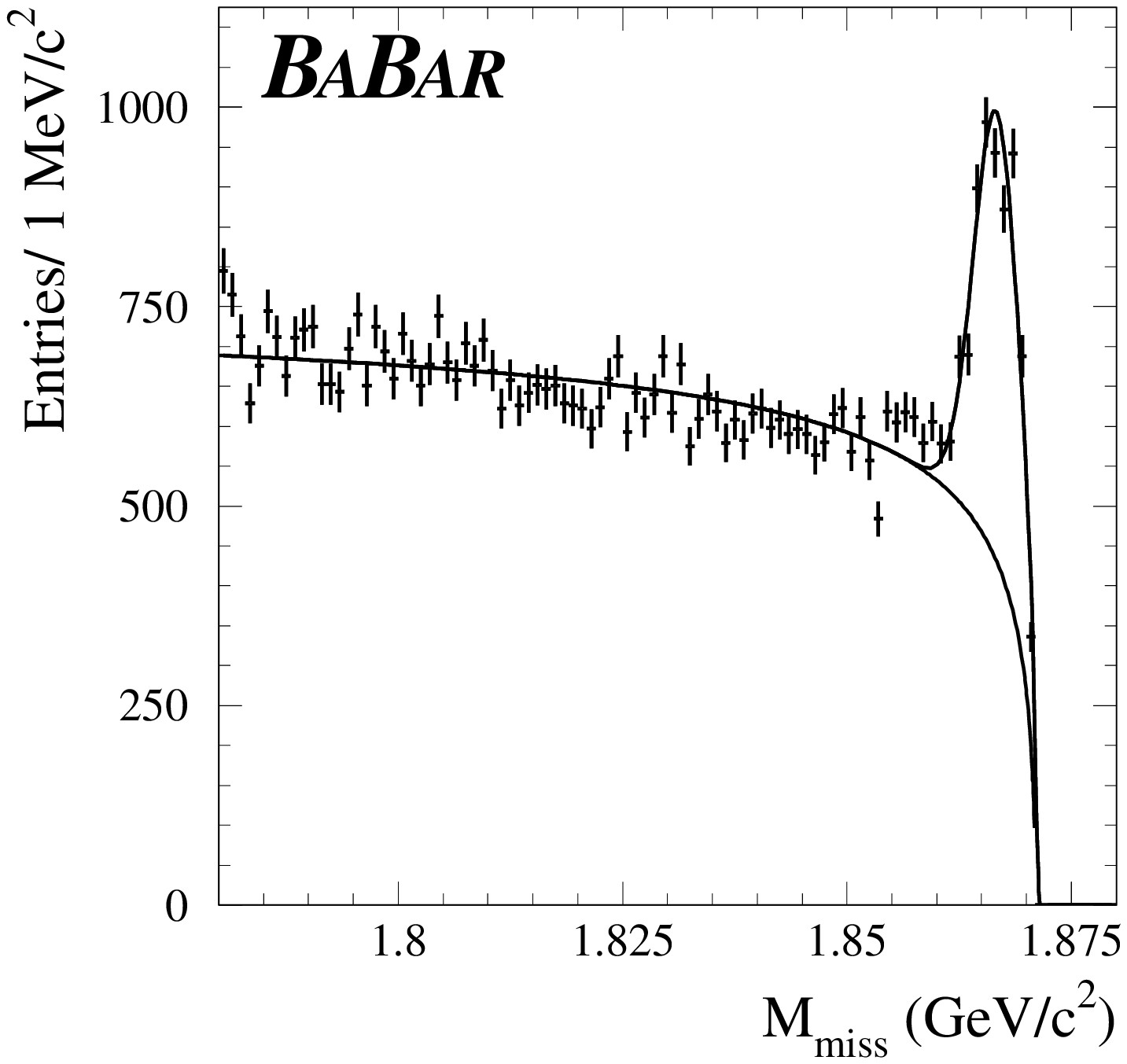}
\hfill
        \includegraphics[width=0.47\textwidth]{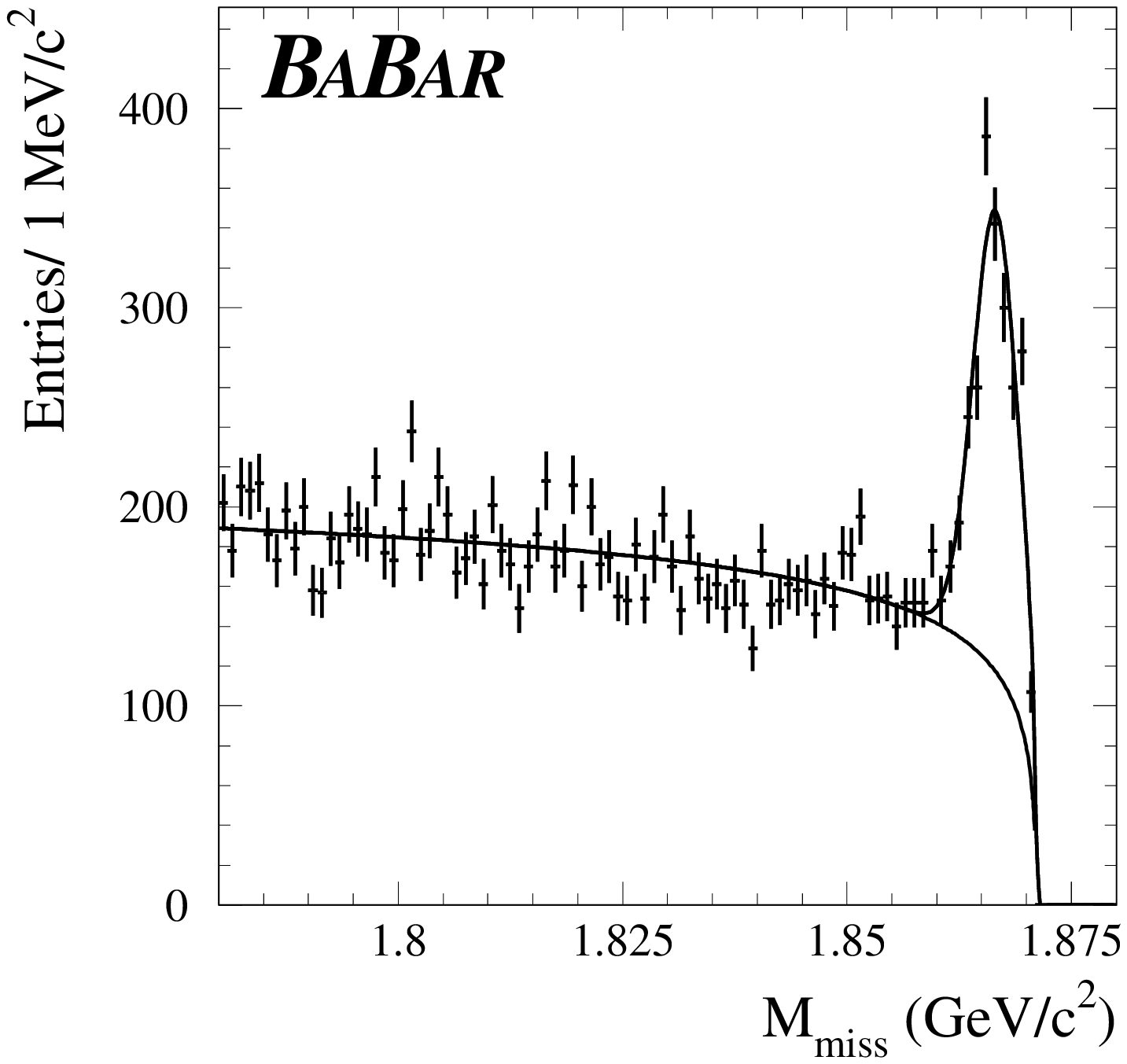}
\\
\parbox[t]{0.47\textwidth}{\caption{$\Dspi$ missing mass distributions
					of data events.  All the \Ds
					decay modes have been combined
					in this plot. The curves show
					the result of the fit (see
					text) indicating the signal
					and background distributions.}
					\label{fig:DsDst_sum}} \hfill
\parbox[t]{0.47\textwidth}{\caption{$\Dsstarpi$ missing mass
					distributions of data events.
					All the \Ds decay modes have
					been combined in this
					plot. The curves show the
					result of the fit (see text)
					indicating the signal and
					background distributions.}
					\label{fig:DsstDst_sum}}
\end{figure}

The missing mass distributions of partially reconstructed \BDstarDss
decays are shown in Fig.~\ref{fig:signal}.  A clear signal peak is observed
in all modes. We perform a binned maximum likelihood fit of these
distributions. The fit function is the sum of a Gaussian distribution
and a background function $f_B$ given by
\begin{equation}
\label{math:bgr}
\begin{array}{c}
f_B(\mmiss) = \frac{C_1\bigl(M_0-\mmiss\bigr)^{C_2}}{C_3+\bigl(M_0-\mmiss\bigr)^{C_2}}
\;,
\end{array}
\end{equation}
where $C_i$ are parameters determined by the fit, and $M_0 = { M}_{
D^*}-{ M}_{\pi} = 1.871$\gevcc is the kinematic end point. The fits
find $3704\pm232$ and $1493\pm 95$ peaking events under the Gaussian peak in the sum of the
$\Dsp\pi^-$ and $\Dspstar\pi^-$ plots, respectively
(Figs.~\ref{fig:DsDst_sum},\ref{fig:DsstDst_sum}). However, due to the
presence of peaking backgrounds, discussed below, further calculation
is needed in order to extract the signal yields and the branching
fractions.

\subsection{Background Study}
\label{sec:bgd-study}
\begin{figure}
\begin{center}
        \includegraphics[width=0.75\textwidth]{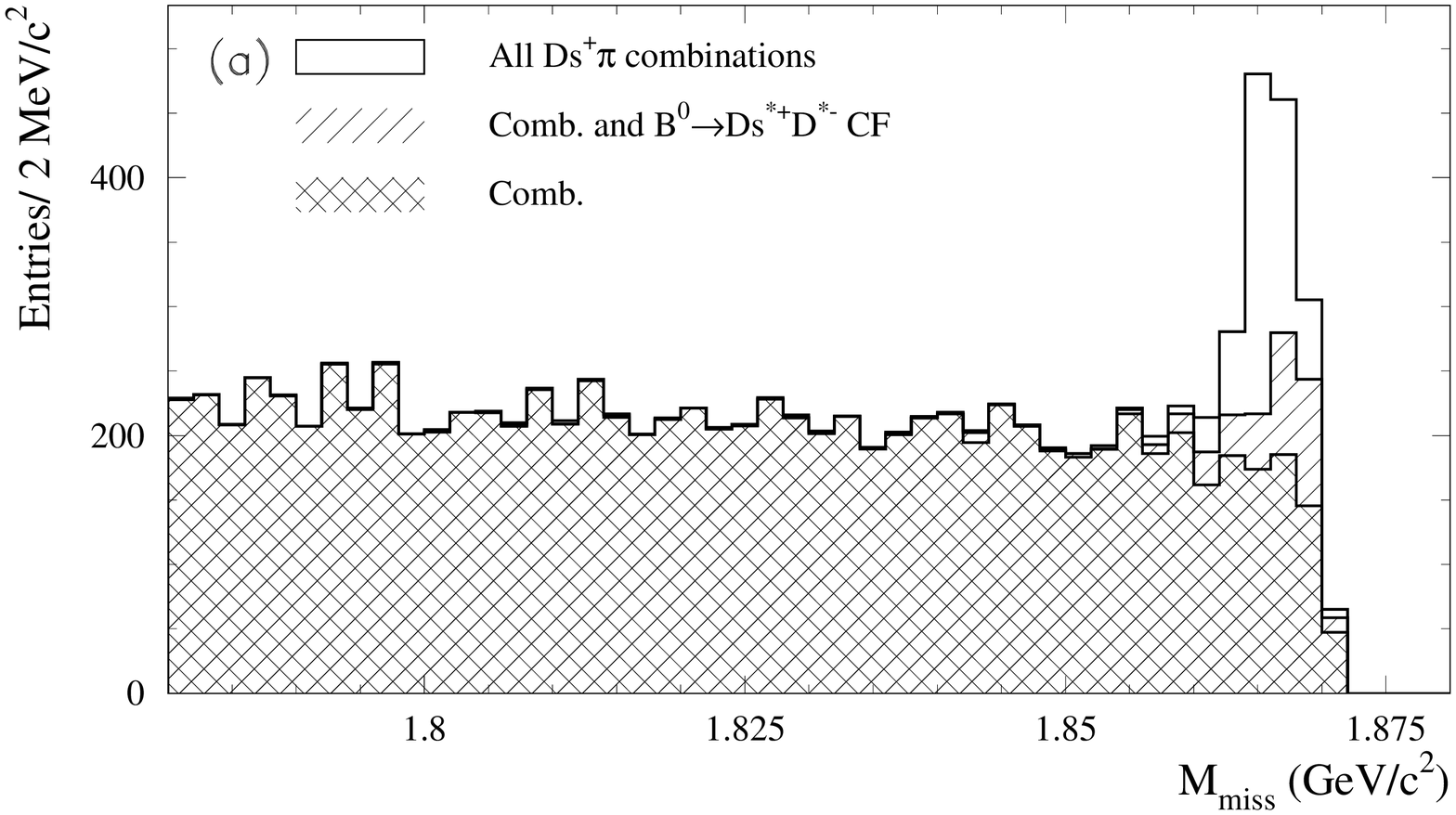} 
        \includegraphics[width=0.75\textwidth]{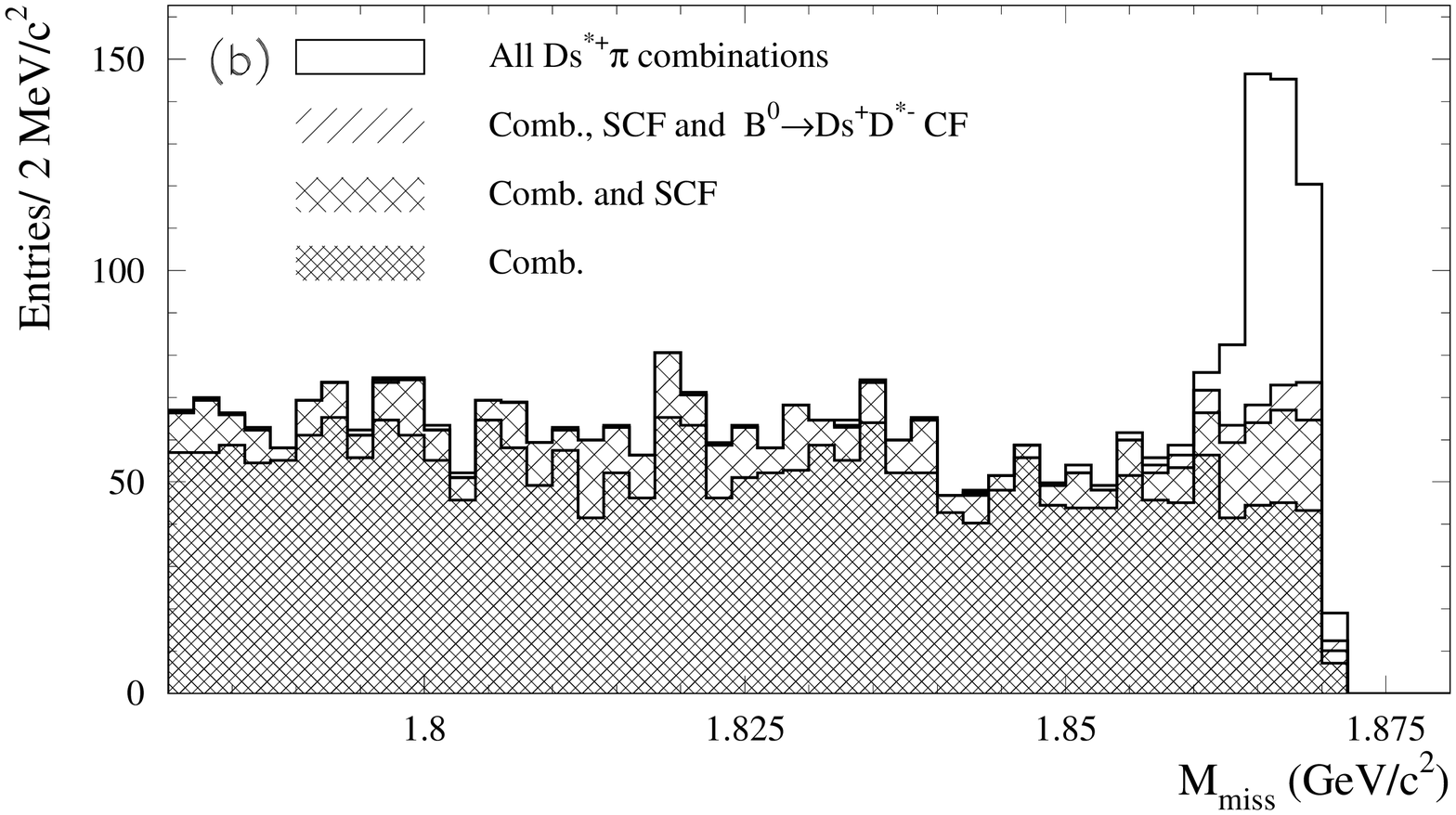} 
	\caption{The missing mass distribution of  
	  (a) $\Dspi$ and (b) $\Dsstarpi$
		Monte Carlo. 
From top to bottom, the overlaid histograms show the contributions of
signal, cross feed (CF), self-cross feed (SCF) ( only for $\Dsstarpi$) and
combinatorial background.
		}
	\label{fig:mc_peak}
\end{center}
\end{figure}

We use a Monte Carlo simulation, which includes both \BB and
$q\bar{q}$ continuum events, to study the missing mass distributions
of the backgrounds.  We consider two kinds of backgrounds: ``peaking''
background is enhanced under the signal peak at the high end of the
missing mass spectrum, and ``non-peaking'' background has a more
uniform missing mass distribution. There are two sources of the peaking
background:
\begin{itemize}
\item
{\bf Cross Feed (CF):} If the soft photon from \Dsgamma decay is not
reconstructed, \BDstarDsstar decays may lead to an enhancement under
the signal peak of the \Dspi missing mass spectrum.
Similarly, the \BDstarDs decays may lead to a peaking enhancement in
the \Dsstarpi \mmiss spectrum, due to the combination of a \Ds with a random
photon.

\item
{\bf Self-Cross Feed (SCF):} This is due to true \BDstarDsstar decays
in which the \Ds is correctly reconstructed, but combined with a
random photon to produce the wrong \Dspstar candidate, resulting in a
peaking enhancement in the \Dsstarpi spectrum.
\end{itemize}
Figure~\ref{fig:mc_peak} illustrates the missing mass distributions
obtained from the Monte Carlo simulation, where the cross feed and the
self-cross feed are shown separately.
Table~\ref{tab:eff} presents the reconstruction efficiency of
correctly reconstructed signal \BDstarDss decays, as well as cross feed and
self-cross feed, for events in the signal region $\mmiss > 1.86$\gevcc.

\begin{table}
\begin{center}
\caption{ The efficiencies of the partially reconstructed \BDstarDss
decays. Columns show the contribution of the different generated modes
to the \Dspi and \Dsstarpi missing mass distributions in the signal
region $\mmiss > 1.86$\gevcc. Two different \BDstarDsstar Monte Carlo
samples have been used, one with longitudinal (long.) and the
other with transverse (transv.) polarization.}
\label{tab:eff}
\begin{tabular}{lcc} \hline
& \multicolumn{2}{c}{Reconstructed mode} \\ 
Generated mode                               &  $\Dspi$          & $\Dsstarpi$       	\\ \hline\hline
\BDstarDs			&  23.6$\pm$1.0\%	& 1.7 $\pm$0.3\%	\\		\hline
\BDstarDsstar (long.)		&  9.0$\pm$0.3\%	& 7.4 $\pm$0.3\%	\\
Self-Cross  Feed		&			& 1.6 $\pm$0.1\%	\\		\hline
\BDstarDsstar (transv.)		&  10.4$\pm$0.3\%	& 6.9 $\pm$0.3\%	\\ 
Self-Cross  Feed		&			& 1.4 $\pm$0.1\%	\\ 	\hline	\hline	
\end{tabular}
\end{center}
\end{table}

In addition to the above backgrounds, we also considered a possible
contribution from the charged and neutral \B decays $\B\to\Dsps
\Dbar^{**}$. These backgrounds were simulated with four $\Dbar^{**}$
states: $\Dstarb_0(j=1/2)$, $\Dbar_1(2420)$, $\Dbar_1(j=1/2)$ and
$\Dstarb_2(2460)$, and their contribution has been determined to be
negligible, due mainly to the \Dsps CM momentum cut.

\begin{figure}
\begin{center}
        \includegraphics[width=0.75\textwidth]{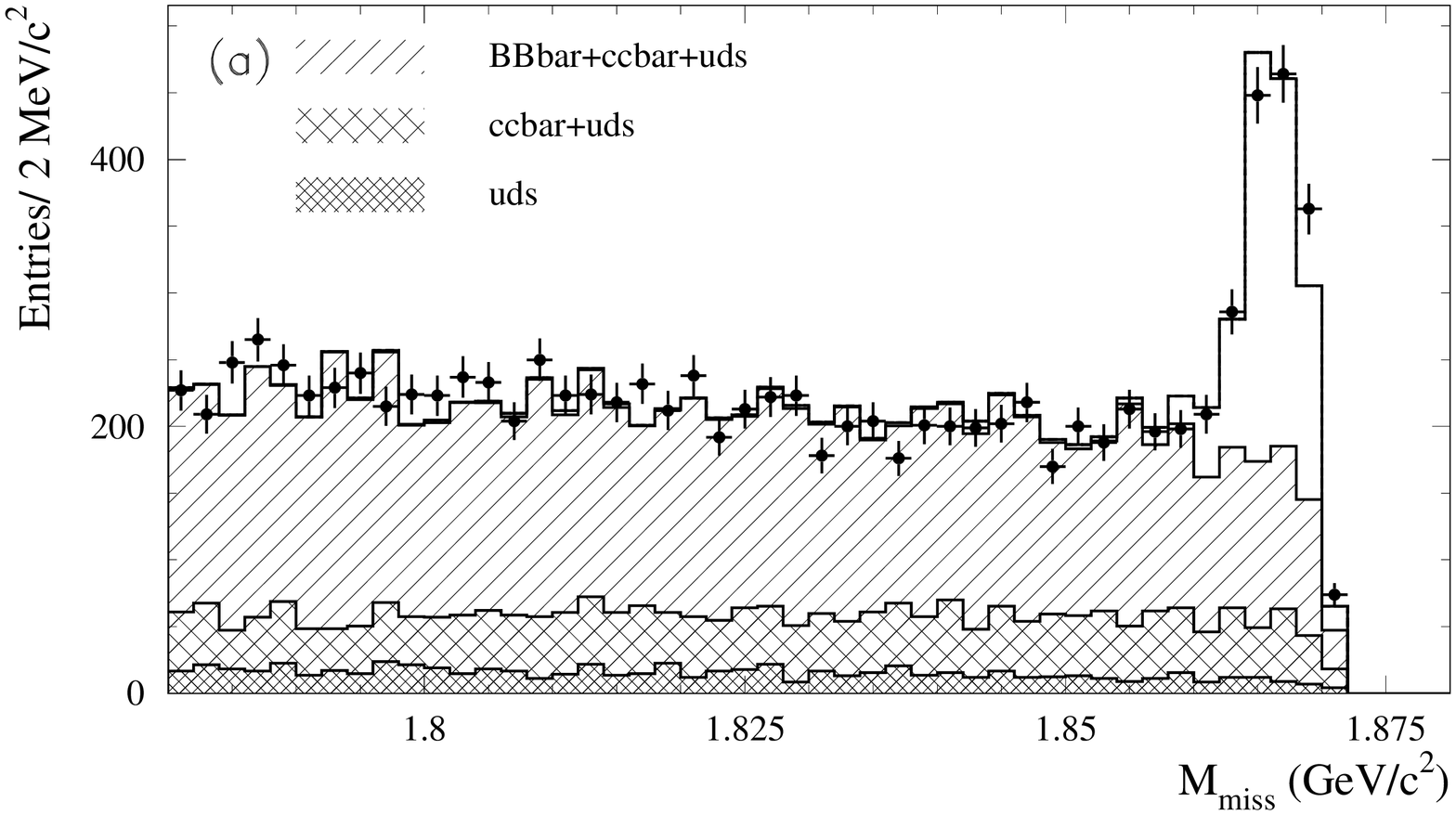} 
        \includegraphics[width=0.75\textwidth]{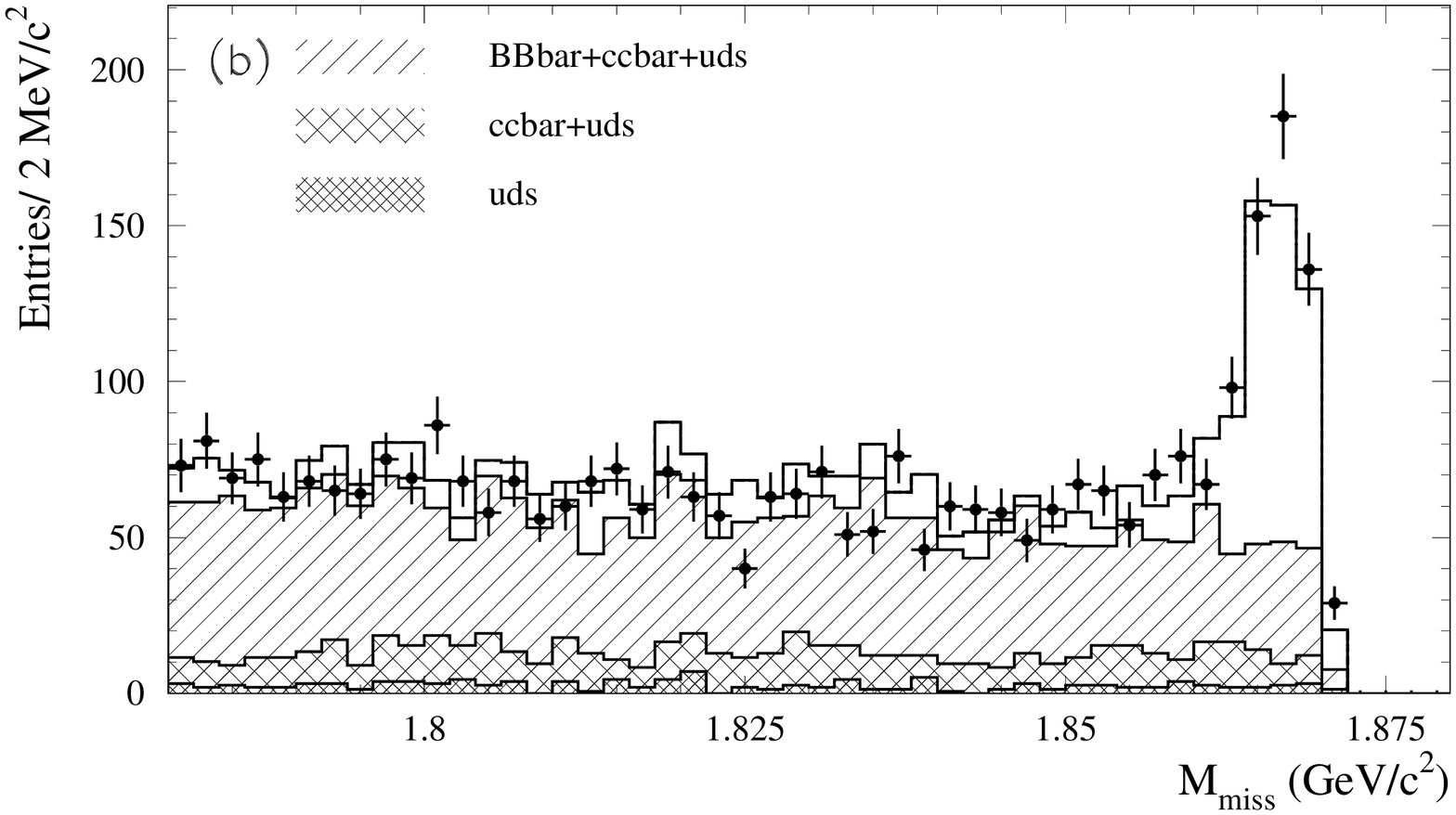} 
	\caption{The missing mass distribution of (a) $\Dspi$ and
		(b) $\Dsstarpi$ combinations
		for data (data points) and Monte Carlo (histogram). 
		The contributions from the $\B\bar{\B}$, \ccbar and $uds$ 
		are shown separately. The CF and SCF backgrounds are included
		in the total histogram, not in the hatched $\BB$ histogram.
		}
	\label{fig:mc}
\end{center}
\end{figure}

Figure~\ref{fig:mc} shows a comparison of the missing mass distributions
in data and Monte Carlo events. We assume 1.05\% and 1.59\% branching fractions 
for the \BDstarDs and \BDstarDsstar decays, respectively, in the Monte Carlo simulation.

\subsection{Branching Fractions}

The number of events in the $\Dspi$ and $\Dsstarpi$ \mmiss peaks is
obtained from the fits described in section \ref{sec:results}. 
In calculating the branching fractions from these yields, we take into
account the fact that the peaks consist not only of correctly
reconstructed signal, but also of cross feed and self-cross feed. This
is done by inverting the $2 \times 2 $ efficiency matrix, whose
diagonal elements correspond to the sum of signal and self-cross feed
efficiencies presented in Table~\ref{tab:eff}, and whose off-diagonal
terms are the cross-feed efficiencies.
The efficiencies corresponding to transverse and longitudinal
polarization of \BDstarDsstar have been weighted according to the
measured polarization (see section~\ref{sec:polarization}).
With this procedure, the \BDstarDss branching fractions are determined
to be
\begin{eqnarray}
\label{math:DsDst}
{\mathcal{B}} (\BDstarDs)  & = & 
(1.03 \pm 0.14 ~\stat \pm 0.13 ~ \syst  \pm 0.26~ \systphipi ) \% \;, \\*[0.2 cm]
{\mathcal{B}} (\BDstarDsstar)  & = & 
(1.97 \pm 0.15 ~ \stat \pm 0.30 ~ \syst  \pm 0.49~ \systphipi  )\%   
\end{eqnarray}
and their sum is 
\begin{equation}
{\Sigma\mathcal{B}} (\BDstarDss)  = 
(3.00 \pm 0.19 ~ \stat  \pm 0.39 ~ \syst  \pm 0.75 ~\systphipi)\%  ,
\end{equation}
where the first error is statistical, the second is
the systematic error from all sources other than the uncertainty in
the \Dsphipi branching fraction, and the third error, which is 
dominant, is due the uncertainty in the \Dsphipi branching fraction
$\brphipi = (3.6\pm 0.9$)\%~\cite{ref:pdg}.
The sources of the systematic error are discussed in section
\ref{sec:Systematics}.

\subsection{Polarization}
\label{sec:polarization}

The measurement of the fraction of the longitudinal polarization
$\Gamma_L/ \Gamma$ in the \BDstarDsstar decay mode is performed using the
events reconstructed in this mode in the signal region
($M_{miss}>1.86$\gevcc).
To reduce the systematic error due to large backgrounds, the
polarization measurement is done with only the channel \Dsphipi, which
has the best signal to background ratio. Two angles are used:
the helicity angle $\theta_{\gamma}$ between the $D^{*-}$ and
the soft photon direction in the \Dspstar rest frame, and 
the helicity angle $\theta_{\pi}$ between the \Dspstar and 
the soft pion direction in the $D^{*-}$ rest frame.
Since the \B meson is not fully reconstructed, we compute
$\theta_{\gamma}$ and $\theta_{\pi}$ by constraining \mmiss to the
nominal $\Dz$ mass~\cite{ref:pdg} to obtain a unique kinematical 
solution for the azimuth $\phi$.

\begin{figure}
	\includegraphics[width=0.9\textwidth]{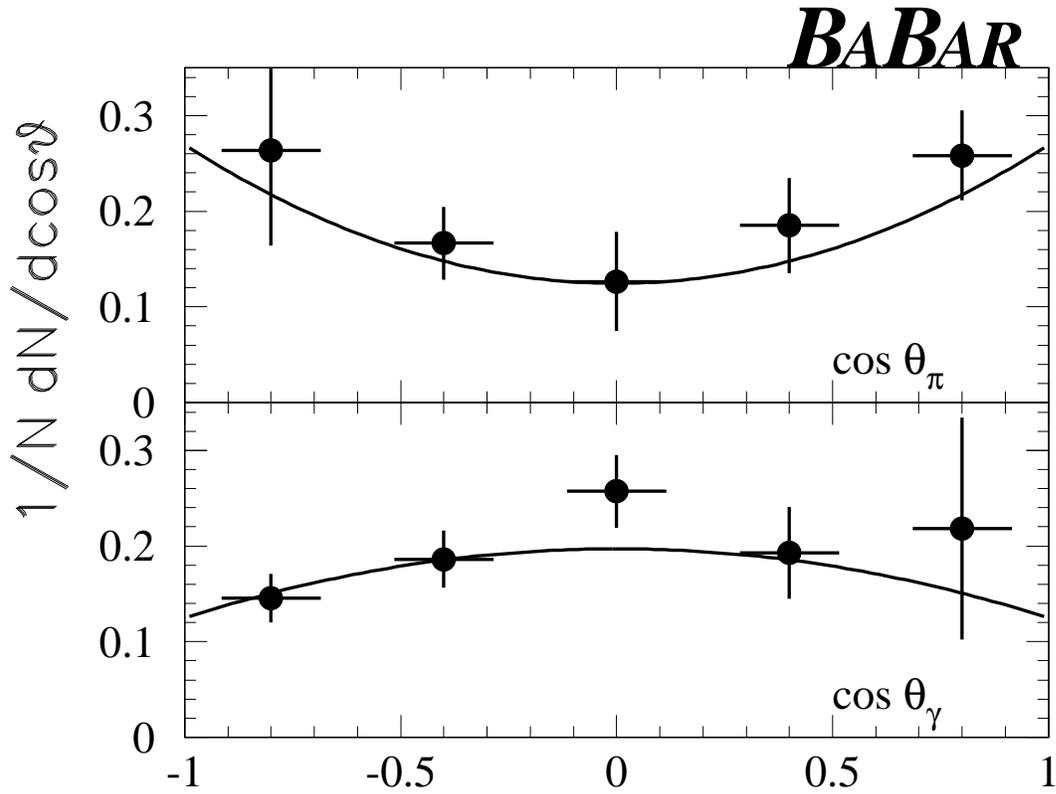}
\caption{Projections of the number of background-subtracted data events
on the $\cos\theta_\gamma$ and $\cos\theta_\pi$ axes.  The result of
the two-dimensional fit is overlaid.}
\label{fig:pol_proj}
\end{figure}

The two dimensional distribution ($\cos\theta_\gamma$,
$\cos\theta_\pi$) is divided in five bins in each dimension.
The combinatorial background, as well as the cross feed and the self-cross 
feed obtained using the Monte Carlo simulation, are subtracted
from this two-dimensional data distribution. The resulting signal
distribution is corrected bin-by-bin for the detector efficiency, 
which is obtained from the simulation separately for each bin.
A two-dimensional binned minimum-$\chi^2$ fit is then performed on the
efficiency-corrected signal distribution using the fit
function
\begin{equation}
\frac{d^2\Gamma}{d\cos\theta_\pi\ d\cos\theta_\gamma }\propto \ \  
\frac{\Gamma_L}{\Gamma}\ \cos^2\theta_\pi \sin^2\theta_\gamma+ 
(1\ -\ \frac{\Gamma_L}{\Gamma}\ \ )  
	\sin^2\theta_\pi\frac{1+\cos^2\theta_\gamma}{4}.
\end{equation}
The resulting fit has a $\chi^2$ of 23.1 for 25 bins with two floating
parameters ($\Gamma_L / \Gamma$ and total
normalization). Fig~\ref{fig:pol_proj} shows the data and the result
of the fit projected on the $\cos\theta_\gamma$ and $\cos\theta_\pi$
axes.

From the fit, the fraction of a longitudinal polarization is determined to be
\begin{equation}
\label{math:GL}
\Gamma_L/\Gamma = (51.9 \pm 5.0   \pm 2.8 )  \%,
\end{equation}
where the first error is statistical and the second is systematic.

\subsection{Systematic Errors}
\label{sec:Systematics}

\begin{table}[!htb]
\caption{Sources of systematic error (\%) for \BDstarDss branching fractions and \BDstarDsstar polarization.}
\begin{center}
\begin{tabular}{lccc}
\hline
\hline
Source					&	\BDstarDs\hspace{0.5cm}		&	\BDstarDsstar	&	$\sigma(\Gamma_L/\Gamma)$	\\ \hline
Background subtraction                          &       2.7    		&	5.9	&	0.5	\\
Monte Carlo statistics                          &       4.2    		&	6.0	&	2.7	\\
Polarization uncertainty                        &       0.8		&	0.5    	&	-	\\ 
Cross Feed					&	3.2		&      	2.4	&	-	\\ 
N$_{B\overline{B}}$                             &       1.6    		&	1.6	&		-\\
$\BR(\phi\rightarrow { K^+K^-})$ 		&       1.6    		&	1.6	&		-\\
Particle identification                         &       1.0    		&	1.0	&	0.1	\\ 
Tracking efficiency                             &       3.6    		&	3.6	&	0.5	\\ 
Soft pion efficiency				&	2.0    		&	2.0	&	0.2	\\ 
Relative branching fractions			&       10.2		&	10.2	&	-	\\
$\BR(\Dsgamma)$  				& -			&	2.7	&		-\\
Photon efficiency                               & -      		&	1.3     &	0.1	\\
$\pi^0$ veto                                    & -      		&       2.7     &	0.3	\\ \hline 
Total systematic error				&  13.1    		&	15.1	& 	2.8	\\ \hline \hline 
\end{tabular}
\end{center}
\label{tab:syst}
\end{table}

The various contributions to the systematic errors of the branching fractions
and polarization measurement are summarized in Table~\ref{tab:syst}.  
The dominant systematic error is due to the uncertainty in our
knowledge of the three \Ds decay branching fractions.
To evaluate the uncertainty due to the background subtraction, the
signal yield is determined in an alternative way, by counting the
number of events in the histogram after a bin-by-bin subtraction of
the background, determined from the Monte Carlo simulation. The
difference of the signal yields obtained in this way from the
results of section~\ref{sec:results} is taken as a systematic error.
This also accounts for the systematic error due to a possible deviation
of the signal shape from a Gaussian.

The Monte Carlo statistical errors in the determination of the signal
and the cross feed efficiencies are propagated to the systematic error.
The uncertainty in the calculation of the \BDstarDsstar polarization is 
propagated to the branching fraction systematic error.
The systematic error due to charged particle reconstruction efficiency
error is 1.2\% times the number of charged particles in the decay. An
additional error of 1.6\% is added in quadrature to account for the
uncertainty in the reconstruction efficiency of the soft pion.

The systematic error due to the excluding \piz overlap (\piz veto) requirement 
was studied by measuring the relative yield of inclusive \Dspstar production in 
data and Monte Carlo events. To evaluate this error, the selection with and without 
the \piz veto was applied for the final photon from \Dsgamma decay.

For the polarization measurement, the level of the various backgrounds
depends on the charged, neutral and particle identification
efficiencies. The fit was repeated varying the background according to
the errors in these efficiencies, and the resulting variations in
$\Gamma_L / \Gamma$ were taken as the associated systematic error.

To check that the simulation accurately reproduces the background \mmiss
distributions in the data, a systematic data-Monte Carlo comparison is
made in control samples containing no signal events. These samples are 
events with $1.78 < \mmiss < 1.85$\gevcc;
events in the \Dsp sideband $ 1.89 < M_{D_s} < 1.95$\gevcc or
$ 1.985 < M_{D_s} < 2.05$\gevcc;
events in the \Dspstar sideband $170 < \Delta M < 300$\mevcc;
wrong sign $\DsspiWS$ combinations in either the $M_{D_s}$ and $\Delta M$
sidebands or signal regions (see section~\ref{sec:selDsps});
and candidates in which \mmiss was calculated using the negative of the 
CM \Dsps momentum $p_{\Dsps}^*$.
The comparison between the data and the Monte Carlo simulation of
these control samples is shown in Table~\ref{tab:mc_ratio}. The
discrepancies indicated in Table~\ref{tab:mc_ratio} are taken into 
account in the calculation of the systematic errors.

\begin{table}[hbt]
\caption{The average value $\langle(N_{\rm D} - N_{\rm MC}) / N_{\rm MC}\rangle$,
averaged over all bins, where $N_{\rm D}$ ($N_{\rm MC}$) is the
number of data (Monte Carlo) events in a given bin of the \mmiss
distribution of the given control sample. SB (SR) refers to the
$M_{D_s}$ or $\Delta M$ sideband (signal region) control sample.  WS
indicates wrong sign $\DsspiWS$ combinations, and $-p_{\Dsps}^*$
indicates that \mmiss was calculated using the negative of the \Dsps
CM momentum. The range of the missing mass $1.78 < \mmiss < 1.87$\gevcc 
of the control sample is used except for the first line.}
\begin{center}
\begin{tabular}{lcc} \hline \hline
Sample type				& 	$\Dspi$		& $\Dsstarpi$	\\ 
\hline \hline
$1.78 < \mmiss < 1.85$\gevcc	& $-0.009\pm0.007$		& $\phantom{-}0.075\pm0.014$	\\ 
SB				& $-0.075\pm0.006$ 		& $\phantom{-}0.007\pm0.022$	\\
SR, WS	 			& $\phantom{-}0.006\pm0.008$	& $\phantom{-}0.044\pm0.015$	\\
SB, WS 				& $-0.060\pm0.007$		& $-0.008\pm0.024$	\\
SR, $-p_{\Dsps}^*$    		& $\phantom{-}0.015\pm0.009$	& $\phantom{-}0.075\pm0.016$	\\
SB, $-p_{\Dsps}^*$		& $-0.062\pm0.007$		& $-0.123\pm0.022$	\\ 
\hline
\hline
Average				& $-0.038\pm0.003$		& $\phantom{-}0.032\pm0.007$	\\
\hline \hline
\end{tabular}
\label{tab:mc_ratio}
\end{center}
\end{table}

\section{Summary}
\label{sec:Summary}

\par
In summary, using the partial reconstruction technique, 
we have measured the branching fractions
$$\BR(\BDstarDs) = (1.03 \pm 0.14 ~ \stat \pm 0.13 ~ \syst \pm 0.26
~\systphipi )\%$$ 
and
$$\BR(\BDstarDsstar) = (1.97 \pm 0.15 ~ \stat \pm 0.30 ~ \syst \pm
0.49 ~ \systphipi)\%.$$ 
The fraction of the longitudinal \Dspstar polarization in
\BDstarDsstar is determined to be
$$ \Gamma_L/\Gamma = (51.9 \pm 5.0 ~ \stat  \pm 2.8 ~ \syst) \% .$$ 
This measurement is consistent with the theoretical prediction of
(53.5$\pm$3.3)\%~\cite{ref:th_factor} assuming factorization. Our preliminary
results are also in a good agreement with previous experimental
results~\cite{ref:cleo-dds-polar, cleo:dsinc}.

\section{Acknowledgments}
\label{sec:Acknowledgments}

We are grateful for the 
extraordinary contributions of our \pep2\ colleagues in
achieving the excellent luminosity and machine conditions
that have made this work possible.
The success of this project also relies critically on the 
expertise and dedication of the computing organizations that 
support \babar.
The collaborating institutions wish to thank 
SLAC for its support and the kind hospitality extended to them. 
This work is supported by the
US Department of Energy
and National Science Foundation, the
Natural Sciences and Engineering Research Council (Canada),
Institute of High Energy Physics (China), the
Commissariat \`a l'Energie Atomique and
Institut National de Physique Nucl\'eaire et de Physique des Particules
(France), the
Bundesministerium f\"ur Bildung und Forschung and
Deutsche Forschungsgemeinschaft
(Germany), the
Istituto Nazionale di Fisica Nucleare (Italy),
the Research Council of Norway, the
Ministry of Science and Technology of the Russian Federation, and the
Particle Physics and Astronomy Research Council (United Kingdom). 
Individuals have received support from 
the A. P. Sloan Foundation, 
the Research Corporation,
and the Alexander von Humboldt Foundation.


\begin{thebibliography}{99}
\bibitem{CBfunc}
The $\Delta M$ distribution of the $\Ds\gamma$ signal is fitted
with the Crystal Ball function
  \[
  f(x) = N \cdot \left\{
  \begin{array}{ll}
    \exp (-\frac{(x - \overline {x})^2}{2 \; \sigma^2})  &;  
( \overline {x} - x)/\sigma >   \alpha \\
    A \times (B - \frac{x - \overline {x}}{\sigma})^{-n} &;  
( \overline {x} - x)/\sigma \leq \alpha \\
  \end{array} \right.  \]
  where $A \equiv \left ( \frac{n}{|\alpha|} \right ) ^n 
\times \exp(-|\alpha|^2/2)$ and
  $B \equiv \frac{n}{|\alpha|} - |\alpha|$.
  $N$ is a normalization factor, $\overline x$ and $\sigma$ are the peak position and width
  of the Gaussian portion of the function, $\alpha$ is the point at which the 
function changes to the power function and $n$ is the exponent of the power function.
  $A$ and $B$ are defined so that the function 
and its first derivative are continuous at $\alpha$.
  More details can be found in D.\ Antreasyan, Crystal Ball Note 321 (1983).
\bibitem{ref:thresh}
The $\Ds\gamma$ background $\Delta M$ distribution is fitted
with the threshold function
\[
f(\Delta M) = p_1 (\Delta M - p_2)^{p_3} e^{p_4(\Delta M-p_2)},
\]
where the four parameters $p_i$ are free in the fit.
\end{thebibliography}
\end{document}